\documentclass[twocolumn,amsmath,amssymb,pre,aps]{revtex4}   
\usepackage{graphicx}
\usepackage{dcolumn}
\usepackage{bm}
\usepackage{color}

\renewcommand{\vec}[1]{\mathbf{#1}}

\newif\ifgraph

\graphtrue             %

\begin{document}
\title{Structure and diffusion of active-passive binary mixtures in a single-file$\dagger$}

\author{Tanwi Debnath\textsuperscript{a}, Shubhadip Nayak\textsuperscript{b}, Poulami Bag\textsuperscript{b}, Debajyoti Debnath\textsuperscript{b} and Pulak Kumar Ghosh\textsuperscript{b}\footnote[1]{Email: pulak.chem@presiuniv.ac.in\\ $\dagger$ This paper is dedicated to 75th Birth Anniversary of Professor S. P. Bhattacharyya}}

\affiliation{\textit{$^{a}$~Department of Chemistry, University of Calcutta, Kolkata 700009, India}}
\affiliation{\textit{$^{b}$~Department of Chemistry, Presidency University, Kolkata 700073, India}}

\date{\today}

\begin{abstract}
We numerically study structure and dynamics of single files composed of active particles, as well as, active-passive binary mixtures. Our simulation results show that when the persistent length of self-propelled particles is much larger than the average inter-particle separation and the self-propulsion velocity is larger than the thermal velocity, particles in the file exist as clusters of various sizes. Average cluster size and structures of the file are very sensitive to self-propulsion properties, thermal fluctuations and composition of the mixture.   In addition to the variation of  file composition, our study considers two sorts of mixture configurations.  One corresponds to the uniform distribution of active passive throughout the mixture in the single file. In the other configuration, active particles are on one side of the file. For the both configurations, even a little fraction of active particles produces a large impact on the structure and dynamics of the file. 
  
\end{abstract}


\maketitle

\section{Introduction}

Both in natural and biology inspired artificial nano-
devices, diffusion under geometric constraints plays pivotal roles in controlling transports of key molecules/ions. In particular, the most often encountered situation  is the diffusion of interacting particles through very narrow channels where particles cannot pass each other.  Diffusion of interacting particles under such constraints occurs as a single-file. A tag particle in a single-file exhibits anomalous diffusion, where mean square displacement in the long time limits grows as, $\langle\Delta x^2\rangle = \langle [x(t)-x(0)]^2\rangle =  2D_\beta t^\beta$, with the exponent $\beta = 0.5$ for most general cases~\cite{Harris,Heckmann,marchesoni-review,Lebowitz,Levitt}. However, $\beta < 0.5$, as well as, $\beta > 0.5$ had also been reported for a tracer diffusion in a single-file~\cite{Misko1,Misko2}.

Historically, the idea of such non-Fickian diffusion of tag particles in a single-file was first introduced during studying transport of water molecules and ions through very narrow channels in membrane~\cite{Hodgkin}. The theories of single file diffusion had been developed from different directions and revisiting the basic random walk problem under  constraints in different contexts~\cite{Heckmann,marchesoni-review,Lebowitz,Levitt,Richards,Fedders,Alexander,Karger1,Berg,mass}.  

Research on single file diffusion started gaining considerable attention with progress in synthesis of nano-materials, which made tracer diffusion in a single-file experimentally accessible. To be specific, synthetic zeolitic materials form networks of channels with radius little larger than the typical size of small organic molecules. Thus, transport of small molecules, like, methane and ethane  through channel networks of zeolitic materials can be described as a single file diffusion.  Experimental studies in such a system, using pulsed force gradient nuclear magnetic resonance, show that mean square displacement is directly proportional to $t^{1/2}$~\cite{Jobic,Kukla}. The most interesting trait of single-file diffusion, the transition from the initial normal diffusion regime to the long-time limits sub-diffusion regime,  has further been confirmed in the several experimental studies~\cite{Lin,Wei,Lutz}. Moreover, single-file like sub-diffusive behavior is observed in diffusion of a tagged monomer in the active polymer\cite{ Rajarshi1, Rajarshi2}. 

Some recent studies~\cite{Lorenzo,Pritha} explore single-file diffusion of a new class of particles which can propel themselves by extracting kinetic energy from their surrounding medium. Such particles are known as active particles, can be of both natural origin (e.g., motor proteins, bacteria, microtubules, actin etc), as well as synthetic ones. Most common class of artificial active particles is the self-propelled Janus particles, consisting of two distinct hemispheres with different physical or chemical properties. Taking advantage of functional asymmetry, these active particles can induce either thermal or concentration gradients~\cite{Paxton,Gibbs,Howse,Volpe,Jiang,Baraban} to propel themselves. Due to self-propulsion mechanisms, active particles are typical non-equilibrium systems\cite{ourPRL,Bao-Quan} with huge application potential for nano-technology and biomedical sciences.

In this paper, we explore structural  and dynamical properties of a system of interacting  active particles, as well as, active-passive binary mixture with varying composition in a single-file. To examine impact of motile active particles in the diffusion of sluggish passive ones, we consider two different file configurations: (1) all the active particles are uniformly distributed over the chain [see Fig.~1(b)]; (2) Active particles are in the one end of the 1D-chain, and the other end contains passive particles only [see Fig.~1(c)]. Our numerical results show that for both types of configurations, even for small concentrations of the active particles the mixture behaves like a system of interacting active particles. Further, particles in the mixture preferably exist as clusters of various sizes for large persistence length of self-propulsion. Diffusion of tagged particles in the single file enhances monotonically with increasing  fraction of active particles in the mixture. All these features attribute to motility transfer from the energetic active particles to the slowly moving passive ones.

We organize the contents of this paper as follows.
Section 2 introduces the key ingredients of the model for a single file: the inter-particle interaction potential ensuring non-passing constraints, and the particle self-propulsion mechanism and associated parameters. In Sec. 3.1, we explore the structure of an active single file both in the thermal and athermal conditions. File structure of active-passive binary mixtures with varying configuration and composition has been investigated in the Sec. 3.2. Section 4 analyses diffusion of single-files. Finally, in Sec. 5, we summarize our results  and  draw a few relevant concluding remarks considering the future direction of work.   

\begin{figure}
\centering
\includegraphics[width=0.45\textwidth,height=0.215\textwidth]{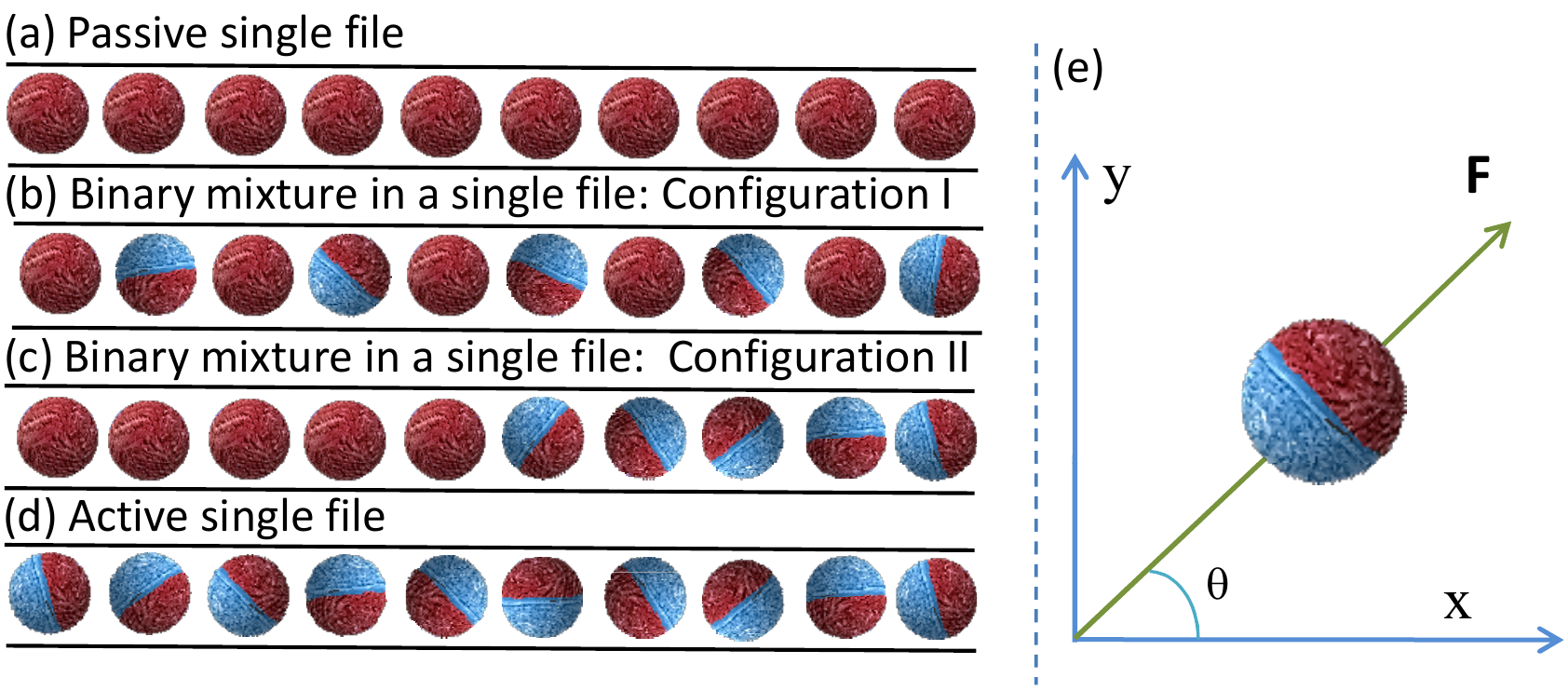}
\caption {(Color online) Schematics for single-file in a 1D narrow channel: (a) Single file of passive particles; (b) {\it Configuration -I}: Single file containing active-passive binary mixture where active particles are uniformly distributed over the entire mixture; (c) {\it Configuration -II}: File contains binary mixture where active particles occupy one half and other half of the file is filled with passive particles;   (d) File comprises active Janus particles;  (e) Schematic of a Janus particle depicting direction of the self-propulsion force $\vec{F}$ (modulus ${F_0}$)  with respect to the channel axis.}
\end{figure}

\section{Model}
We consider a binary mixture 
of $N_a$ active and $N_p$ passive particles in a single file. Fractions of the active and passive particles in the mixture are, $\eta_a = N_a/(N_a+N_p)$ and $\eta_p = N_p/(N_a+N_p)$, respectively. Active particles are characterized by their self-propulsion velocity, $v_0$ and its persistence time, $\tau_\theta$. Further, all particles, $N=N_a+N_p$, are represented by disks of radius $r_0$ and they interact with each other via a truncated Lennard-Jones (L-J) potential. The interaction energy ($V_{ij}$) between a pair of particles separated by a distance $r_{ij}$ is given by, 
\begin{eqnarray}\label{LJ}
V_{ij} &=& 4\varepsilon\left[\left(\frac{\sigma}{r_{ij}}\right)^{12} -\left(\frac{\sigma}{r_{ij}}\right)^{6} \right], \;\; {\rm if}\;\;  r_{ij} \leq r_m \nonumber \\
  &=& 0 \;\; {\rm otherwise},
\end{eqnarray}
where $r_m$ locates the potential minimum and $\sigma$ is the diameter of the particle.
Dynamics of a tagged particle in the single file can be described  by the following set of overdamped Langevin equation,
\begin{eqnarray}\label{LE}
 \gamma \dot{{x}_i}  &=& \sum_{j} {F}_{ij} +   {F}_{0,i} \cos{\theta_i} + \vec{\xi}_{i}^x (t), \nonumber \\
\dot{\theta_i} &=& \,\xi_i^{\theta} (t).
\end{eqnarray}
Where, $\gamma$ is the damping strength depending on the particle's size and viscosity of the medium. The $i-{th}$ particle at the position $x_i$ encounters inter-particle repulsive interaction $\sum_{j}F_{ij}$  [derived from Lennard-Jones potential~(\ref{LJ})] with its neighbouring ones.  This force does not allow particles to pass each other. In addition to these interactions, particle motion is governed by the combined action of self-propulsion and equilibrium thermal fluctuations.

For active particles we consider self-propulsion velocity ($\vec{v}_{0,i}\equiv\vec{F}_{i}/\gamma$) with a constant modulus $v_0$ ($\equiv {F}_{0,i}/\gamma$). On the other hand,   $\vec{v}_{0,i} =0$ for passive particles.  The self-propulsion velocity $\vec{v}_{0,i}$ is oriented at an angle $\theta_i$ with respect to the $x$-axis [as depicted in Fig.~1(e)]. However, the direction ($\theta $) changes   due to rotational diffusion described  by the Wiener process of Eq.(\ref{LE}), where $\langle
\xi_i^{\theta}(t)\rangle=0$ and $\langle
\xi_i^{\theta}(t)\xi_j^{\theta}(t')\rangle=2D_\theta \delta_{ij}\delta (t-t')$. 
The rotational diffusion constant $D_\theta$ is related to the persistence time and length of self-propulsion velocity as, $\tau_\theta = 1/D_\theta$ and $l_\theta = \tau_\theta v_0$, respectively. From the correlation function, $\langle \cos \theta_i (t) \cos {\theta}_i (0)
\rangle = \langle \sin \theta_i (t)\sin \theta_i (0)\rangle =(1/2)\exp[{-D_{\theta}|t|}]$, it is apparent that
$D_\theta$ coincides with the rotational relaxation rate of the
self-propulsion velocity ${\vec v}_0(t)$.

The last term in the Eq.~(\ref{LE}) represents zero mean, delta-correlated, $\langle
\xi_i^{x}(t)\xi_j^{x}(0)\rangle=2D_0\gamma^2 \delta_{ij}\delta(t-t')$ thermal fluctuations with Gaussian distribution. Where, $D_0$  is the free space diffusion constant in the absence of self-propulsion.

For a passive particle, both the rotational diffusion  and translational diffusions are of thermal origin. For a spherical
colloidal particle of radius $r_0$, the rotational and translational diffusion constants in a medium  with viscosity $\eta_v$ and  at the temperature $T$, can be expressed as, $D_{\theta} = k_BT/8\pi\eta_v r_0^3$ and $D_{0} = k_BT/6\pi\eta_v r_0$, respectively. However, for an active particle rotational diffusion can also depend on the mechanisms fueling its
self-propulsion.  For this reason, $v_0$, $D_\theta$ and thermal diffusion constant $D_0$ are treated
here as independent model parameters \cite{Howse, cataly2, Teeffelen}.

The coupled  equations~(\ref{LE}) have been numerically integrated using a standard Milstein algorithm \cite{Kloeden} to explore structure of the active-passive mixture  as well as diffusion of a tagged particle in the single files. Specifically, we estimate the fraction of particles form cluster, cluster size, and the mean square displacement of a tagged particle.  The numerical integration has been performed using a very short time step, $10^{-5} - 10^{-6}$, to ensure numerical stability. At $t=0$, we assume all particles are uniformly distributed in the 1D box of length $L$ with random orientation of self-propelled velocity $\vec{v_0}$.  To keep the mixture density constant with respect to both active and passive species we use periodic boundary conditions.

In our simulation we choose particle diameter $\sigma$ as the unit of length. The unit time and force, respectively, are defined as, $\gamma\sigma^2/\varepsilon$ and $\varepsilon/\sigma$. Throughout this paper, we set $\varepsilon = 1,\;\gamma = 1, {\rm and }\; \sigma = 1$.


\section{Single file Structure}
\begin{figure}
\centering
\includegraphics[width=0.4\textwidth,height=0.45\textwidth]{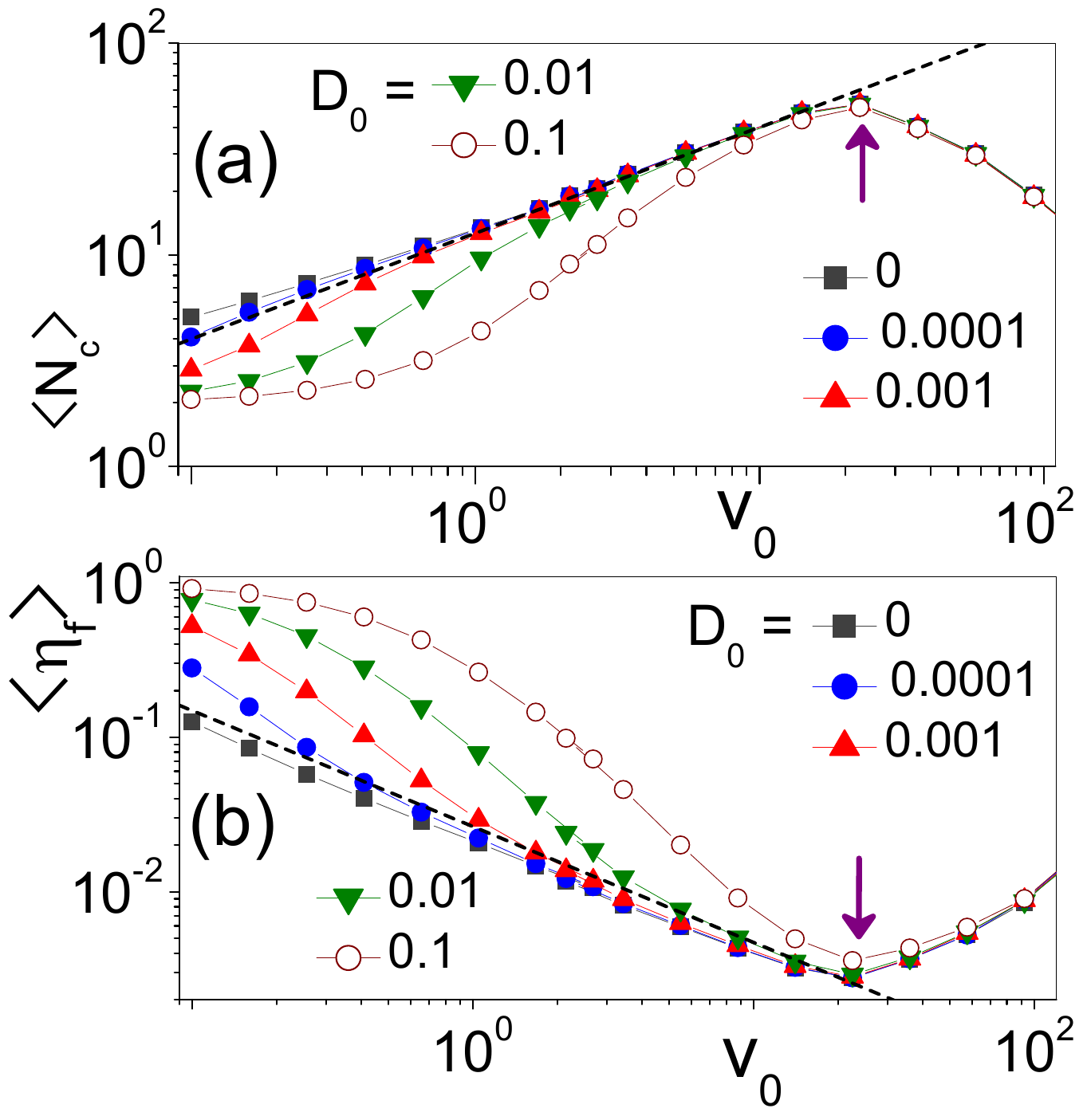}
\caption {(Color online) (a) Average particle number in a cluster $\langle N_c \rangle $ as a function of self-propulsion velocity $v_0$ for different thermal diffusion constant $D_0$ (see legends). (b) Average fraction of particles exist as a single particle ($\langle \eta_{f} \rangle$) in the mixture versus $v_0$ for different $D_0$ as shown in the legends. The dashed lines in (a) and (b) correspond to the Eq.(\ref{cluster-1}) and Eq.(\ref{free}), respectively. The remaining model parameters are (unless reported otherwise in the legends): $D_\theta = 0.002,  \; {\rm and} \; \rho_N = 0.16$.}
\end{figure}

We first explore the structure of  active-passive binary mixture in a single file. Our simulation results reveal that particles in the mixture exist as clusters of various sizes as long as self-propulsion length ($l_\theta$) surpasses the average separation between two adjacent particles. Similar to the earlier results~\cite{MP1,MP2,MP3,MP4,MP5,MP6,MP7,MP8} for higher dimensions, in one dimension the formation of aggregates is solely attributed to the motility of active particles and  interparticle attractive forces  play no role. Such an interesting phenomenon is referred to as motility induced phase separation.

To characterize structure of the mixture we numerically estimate average cluster size, average fraction of particles exist as single and cluster size distributions. We assume two particles form a cluster whenever their centre to center distance  is less than the cut-off distance $r_m$ where particles start encountering repulsive force. Clusters are counted in the long time limits where  effects of transient processes can safely be ignored. Average cluster size, { \it i.e.,} average number of particles in a cluster ($ \langle  N_{c} \rangle$) is estimated by taking average over as many as $1000$ observations. The time lag between two observations is chosen to be much larger than the rotational relaxation time $\tau_\theta$. 
\begin{figure}
\centering
\includegraphics[width=0.4\textwidth,height=0.45\textwidth]{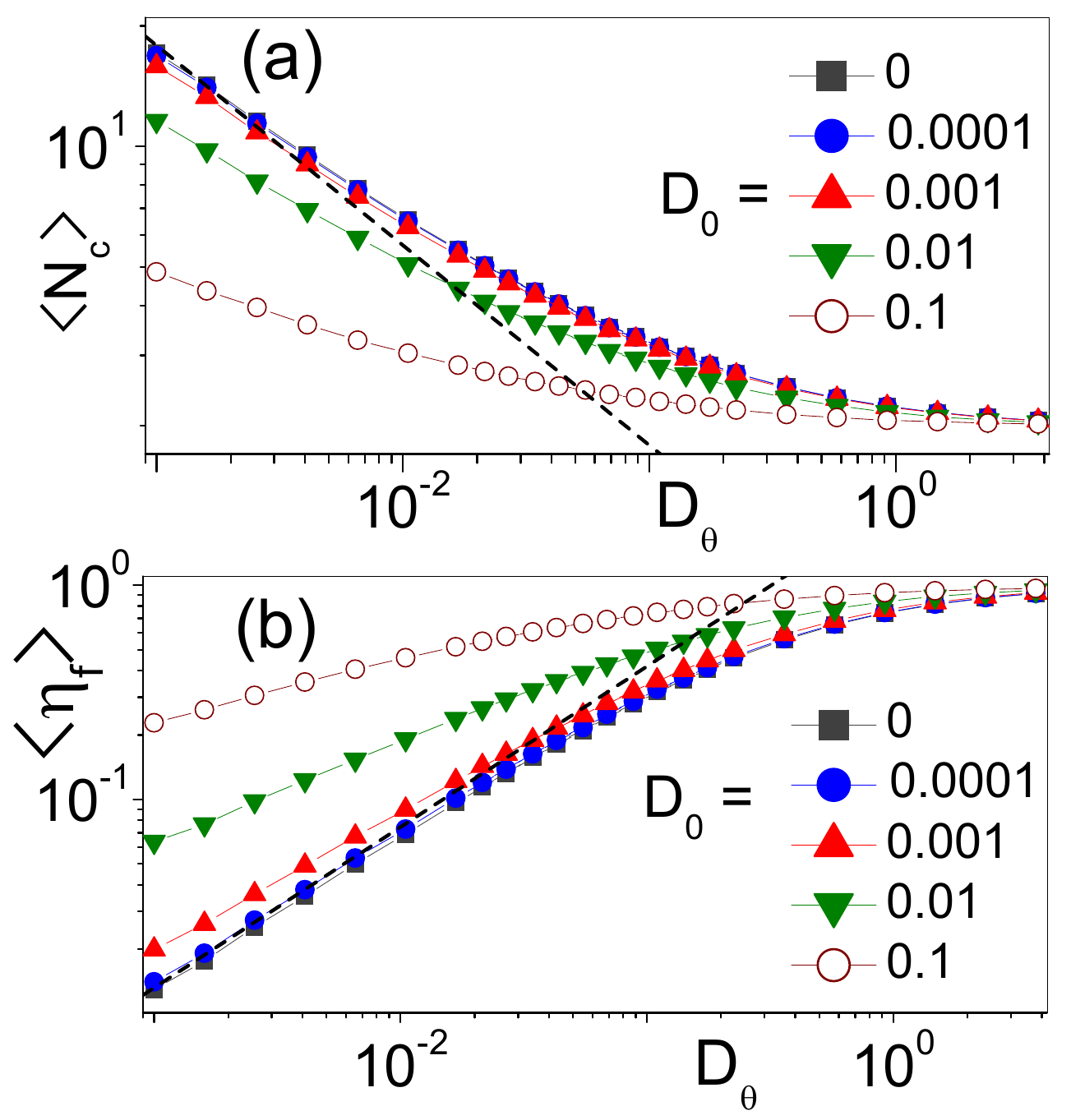}
\caption {(Color online) (a) $\langle N_c \rangle \;  vs.  \; D_\theta $ and (b) $ \langle \eta_f \rangle \;  vs.  \; D_\theta $ for different thermal diffusion constant $D_0$ (see legends). The dashed lines in (a) and (b) correspond to the Eq.(\ref{cluster-1}) and Eq.(\ref{free}), respectively.  The remaining model parameters are (unless reported otherwise in the legends): $ \; v_0 = 1.0, \; {\rm and} \; \rho_N = 0.16$.}
\end{figure}
\subsection{Structure of an active single file} In Fig.~2(a,b) and Fig.~3(a,b) we present the average cluster size and the average fraction of  particles which does not form cluster ($\langle \eta_f \rangle$) as a function of self-propulsion parameters in the limit, $\eta_a \rightarrow 1$. It is apparent from simulation results that thermal noise has a strong impact on the structure of the file. When the self-propulsion contribution in the diffusion is much stronger than thermal diffusion, $v_0^2/2D_\theta \gg D_0$, the average cluster size $\langle N_c \rangle$ grows linearly with the square root of self-propulsion velocity, rotational relaxation time and particle number density. Average cluster size can be expressed in more general empirical relation,  
\begin{eqnarray}\label{cluster-1}
\langle N_{c} \rangle = A \left( v_0 \rho_N \tau_\theta \right)^{\beta_1}  
\end{eqnarray}
where, $\beta_1 \sim 1/2$ in the athermal condition. Further, $\rho_N$ is the particle density and its reciprocal is the average interparticle separation.   This is in accord with the earlier report~\cite{Pritha}. The pre-factor $A$ is a constant and it is close to $\sqrt{2}$ for the best fitting numerical data. Note that Eq.~(\ref{cluster-1}) is valid as long as self-propulsion persistence length is much larger than the average inter particle spacing, $l_\theta \gg 1/\rho_N$ and self-propulsion strength is not too strong.  

\begin{figure}
\centering
\includegraphics[width=0.4\textwidth,height=0.6\textwidth]{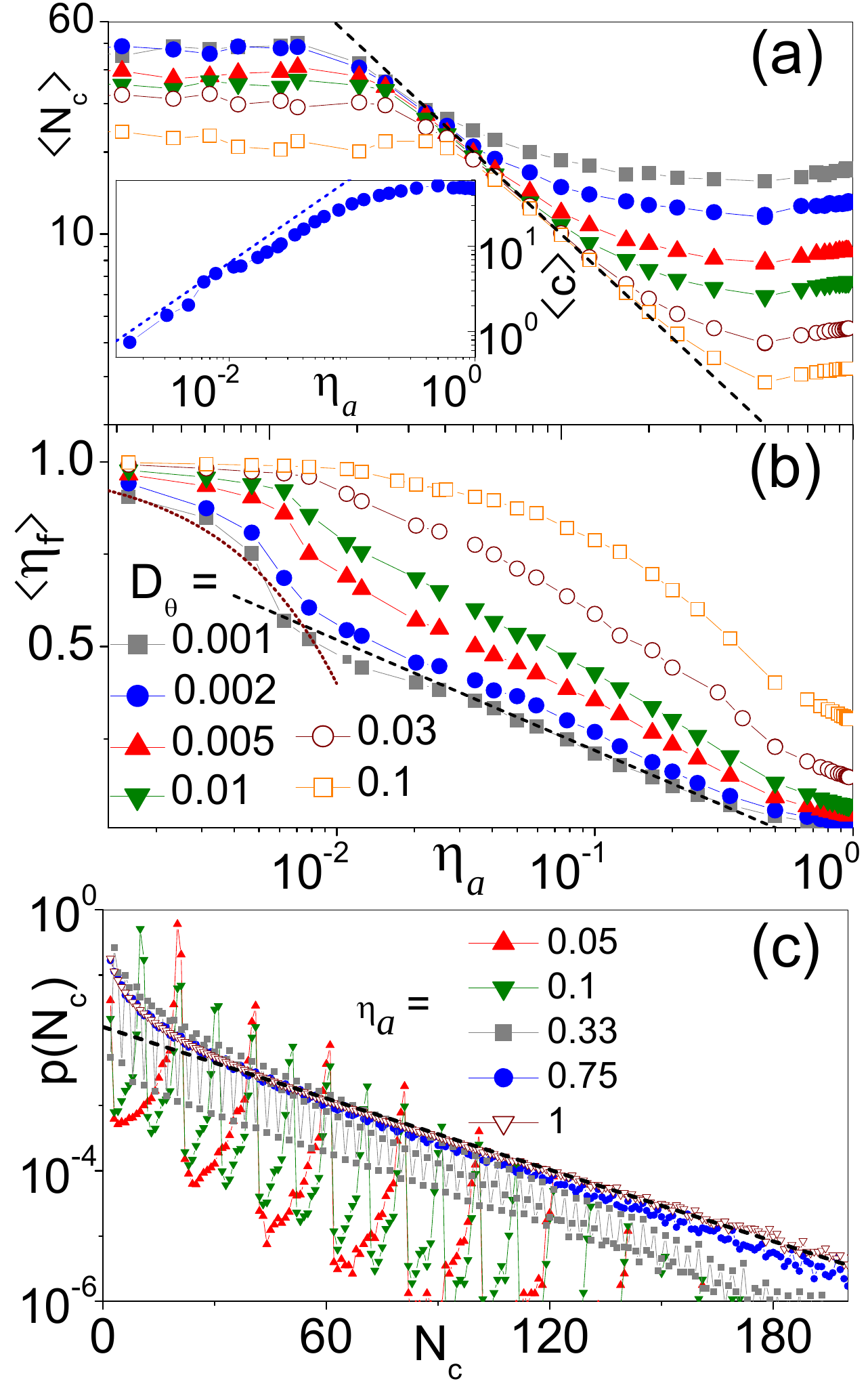}
\caption {(Color online) (a) $\langle N_c \rangle \;  vs.  \; \eta_a$ for different rotational diffusion constant $D_\theta$ (see legends) when active particles are uniformly distributed over the entire length of the file ({\it configuration-I}). The dashed line displays Eq.(\ref{cluster-m1}). Inset depicts average cluster numbers $\langle c \rangle$ as a function of $\eta_a$. Dashed line here indicates that the average cluster number is equal to the number of active particles ($\langle c \rangle = N \eta_a$). (b) Similar figure as (a) but $ \langle \eta_f \rangle \;  vs.  \;\eta_a $.  The dotted line represents Eq.(\ref{free-3}). The dashed line is  a
guide to the eye showing logarithmic decay of the form: $\langle \eta_f \rangle = A_1+B_1 ln(\eta_a)$, where $A_1$ and $B_1$ are functions of self-propulsion parameters. (c) Cluster size distribution with varying mixture composition (see legends). The dashed line is the estimation of exponentially decaying tail based on the Eq.(\ref{cluster-dis1}). The  model parameters are (unless reported otherwise in the legends): $ \; v_0 = 1.0, \; \rho_N = 0.16, D_0 = 0,\; {\rm and} \; D_\theta = 0.002$. }
\end{figure}

Figure 2(a) shows that beyond a certain threshold value of the self-propulsion velocity, the cluster size monotonically decays with $v_0$. This is due to the fact that with growing self-propulsion force, penetration length of the self-propellers increases leading to decrease of the effective particle size. Thus, the density of the system gradually decays with $v_0$. The penetration length can be estimated by equating repulsion force due to L-J potential with the self-propulsion force, $dV/dr + F_0=0$. When two particles touch each other, $|x_i - x_j| = \sigma$, the force balance condition produces the threshold    value of $v_0$,   
\begin{eqnarray}\label{vcut}
v_0^c = \frac{24\epsilon}{\sigma} 
\end{eqnarray}
When,  $v_0 >v_0^c$, particles' self-propulsion force becomes strong enough to penetrate each other overcoming repulsive interaction derived from L-J potential. This leads to a drop of the effective density of the system. Thus, one can expect a change in the trend of variation for $\langle N_c \rangle  \; vs. \; v_0$ and $\langle \eta_f \rangle \; vs. \; v_0$ around $v_0 = v_0^c$.     

Simulation results in Fig.2 well corroborate above reasoning.   $\langle N_c \rangle  \;{\rm versus} \; v_0$ displays a maximum and $\langle \eta_f \rangle  \;{\rm versus} \; v_0$ exhibits a  minimum around  $v_0 = v_0^c$. This estimation is indicated by vertical arrows [see Fig.
2(a, b)].   

Both in the thermal and athermal conditions, $\langle \eta_f \rangle$ monotonically decays with self-propulsion parameters, $v_0$ and $\tau_\theta$, as well as $\rho_N$  when $v_0 < v_0^c$. Our simulation results show that for $D_0 = 0$ and $l_\theta \gg  1/\rho_N $ following empirical relation holds,   
\begin{eqnarray}\label{free}
\langle \eta_f \rangle = B\left(\frac{1}{v_0 \tau_\theta \rho_N } \right)^{\beta_2} 
\end{eqnarray}
Simulation data in Fig.2(b) and Fig.3(b) are best fitted for $B=1/2^{\beta_2}$ and $\beta_2=0.75$. Estimation based on the empirical relation [Eq.~(\ref{free})] is indicated by a dashed line.  As anticipated, when the self-propulsion length is comparable or much smaller than the inter-particle spacing, $l_\theta \sim 1/\rho_N$, the equation ~(\ref{free}) loses its validity. In this limit, the self-propellers change their direction before they travel an appropriate distance to collide and form clusters.  Thus, $\langle N_c \rangle $ and   $\langle \eta_f \rangle $  become insensitive to the particle's rotational dynamics as long as, $D_\theta \gg \rho_N v_0$ .      Further, in this regime, cluster formation becomes a rare event due to frequent flipping of self-propulsion velocity direction. If an aggregate happens to form, there would be overwhelmingly more probability for the two-particle clusters than that of other sizes. Thus, for the vanishingly small rotational relaxation time, $\langle N_c \rangle \rightarrow 2$ and $\langle \eta_f \rangle \rightarrow 1$. 

We conclude this section with notes on the impact of the thermal translational diffusion on, $\langle N_c \rangle $ and $\langle \eta_f \rangle $. Simulation results in Fig.~2  and Fig.~3 show that thermal fluctuations even of very little intensity ($D_0 \ll v^2_0/2D_\theta$, as well as, $v_0 \gg \sqrt{\gamma D_0}$) have noticeable impact on the file structure.  As usual, the thermal noise destroys the order structure, $\langle N_c \rangle$ monotonically decays, whereas, $\langle \eta_f \rangle$ enhances with thermal noise strength. When the  average thermal velocity ($v_{th} = \sqrt{\gamma D_0}$) is comparable or larger than $v_0$, the structure of the file remains insensitive to the self-propulsion parameters. 
\begin{figure}
\centering
\includegraphics[width=0.4\textwidth,height=0.5\textwidth]{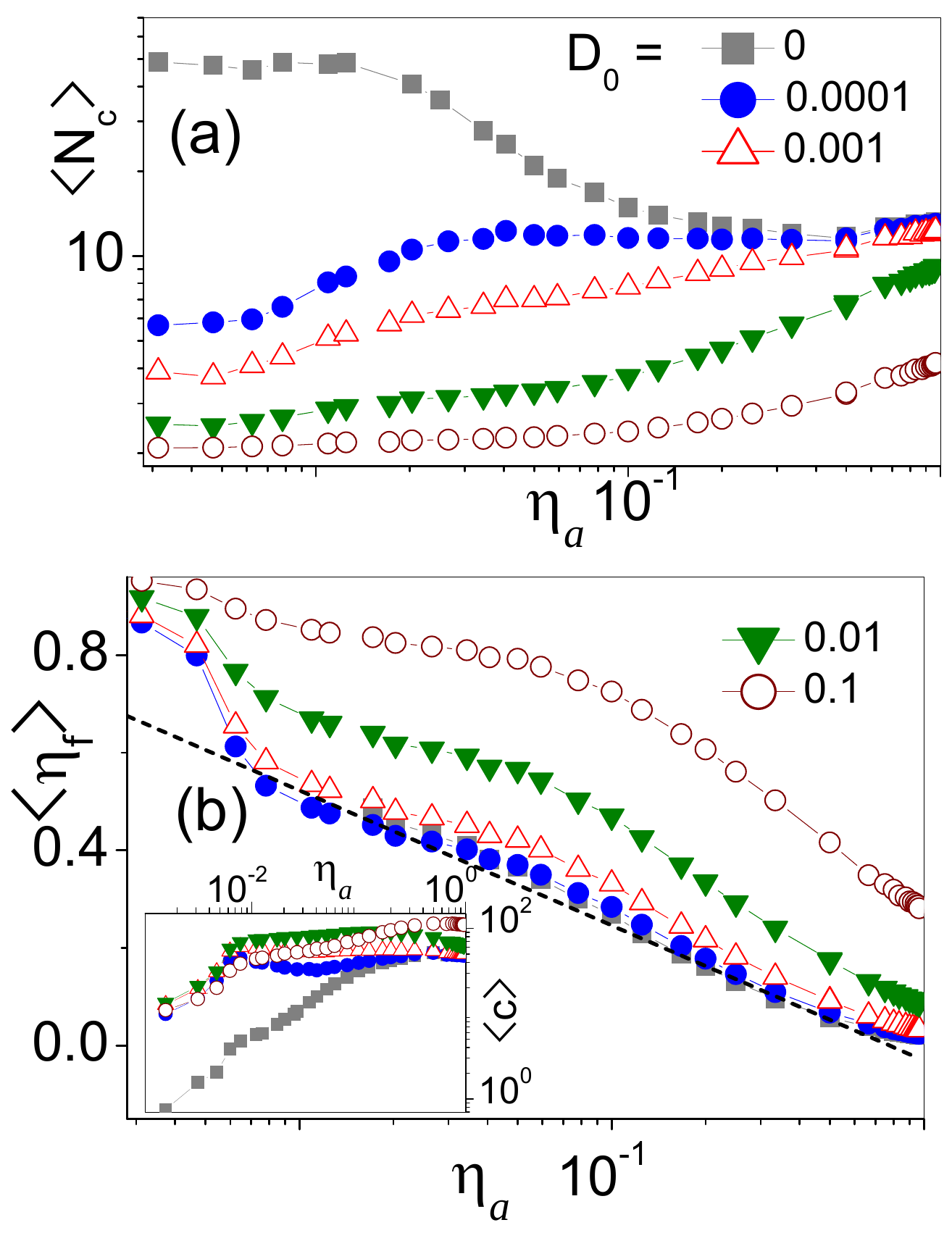}
\caption {(Color online) (a) $\langle N_c \rangle \;  vs.  \; \eta_a$, and (b) $ \langle \eta_f \rangle \;  vs.  \;\eta_a $  for different thermal diffusion constant $D_0$ (see legends) when active species are uniformly distributed over the entire length of the file ({\it configuration-I}). The dashed line in panel (b) guides to the eye showing logarithmic decay similar as in Fig.~4(b). Further, the inset in the panel (b) shows average cluster number versus $\eta_a$. The model parameters are (unless reported otherwise in the legends): $\; v_0 = 1.0, \; \rho_N = 0.16, \; {\rm and} \; D_\theta = 0.002$. }
\end{figure}

\subsection{Structure of binary mixture in a single file }
In a passive single file, particles are almost uniformly distributed with an average interparticle  separation $l_s = 1/\rho_N$. Now, with  gradually introducing active particles the structure of the file changes drastically. Active particles push the 
sluggish passive ones and form aggregates of different sizes. For a systematic analysis, we consider two possible configurations of the binary mixtures. {\it Configuration I}: Active particles are uniformly distributed throughout the mixture [see Fig.1(b)]; {\it Configuration II}: Active particles are in the one side of the file and the other side contains passive particles only (see Fig.1(c)).

For the both configurations, for a vanishingly small fraction of active particles in the mixture very large clusters are formed however they are very few in number. Noticeably, the structure of $\langle N_c \rangle$ versus $\eta_a$,  average cluster number ($\langle c \rangle$) over the length $L$ versus $\eta_a$,  and $\langle \eta_f \rangle$ versus $\eta_a$  are considerably different for two configurations, as well as, they are very sensitive to the thermal fluctuations [see Fig.~(4-7)].

\subsection*{Configuration I:  Active particles are uniformly distributed throughout the mixture}
{\it Athermal regime ---} Here we begin our discussion considering the situation of zero thermal fluctuations ($D = 0$). Figure 4 shows that in the athermal condition, the average cluster number is equal to the number of the active particles in the mixture. It produces, $\langle c \rangle \sim N_a = \eta_a N$ (indicated by the dashed line in the inset of Fig.4(a)). Where, $N_a$ is the number of active particles in the mixture.   However, the cluster size $\langle N_c \rangle$ remains insensitive to $\eta_a$ as long as the dynamics of an active particle is not affected by other active particles. Injected active particles push all the passive ones they encounter in their path and form a cluster.  Suppose, an active particle is injected in a system of passive particles with average separation $l_s$.  Further, we assume that the active particle is oriented along the channel axis at the starting point and it moves with an average velocity $\overline{v}=2v_0/\pi$ until its direction reverts by diffusing over an angle $\pi$. These assumption yields particles in the aggregate,
\begin{eqnarray}\label{cluster-2}
\langle N_c \rangle = \sqrt{\frac{2\pi v_0}{l_s D_\theta}+\frac{1}{4}}-\frac{1}{2}
\end{eqnarray}   
As time elapses the system loses its structural uniformity and cluster size changes drastically. This is due to the fact that even when the active pusher reverts its direction the passive particles remain close to each due to the absence of the thermal fluctuations. Thus, in the next turn when the active particle encounters this closely spaced passive crowd, the cluster would form aggregating more particles. Further, on reverting direction the active particle gets a long free run to reach the passive crowds fast. Therefore, the average cluster size, $\langle N_c \rangle$, in the long time limit is considerably different from the Eq.~(\ref{cluster-2}).

\begin{figure}
\centering
\includegraphics[width=0.4\textwidth,height=0.7\textwidth]{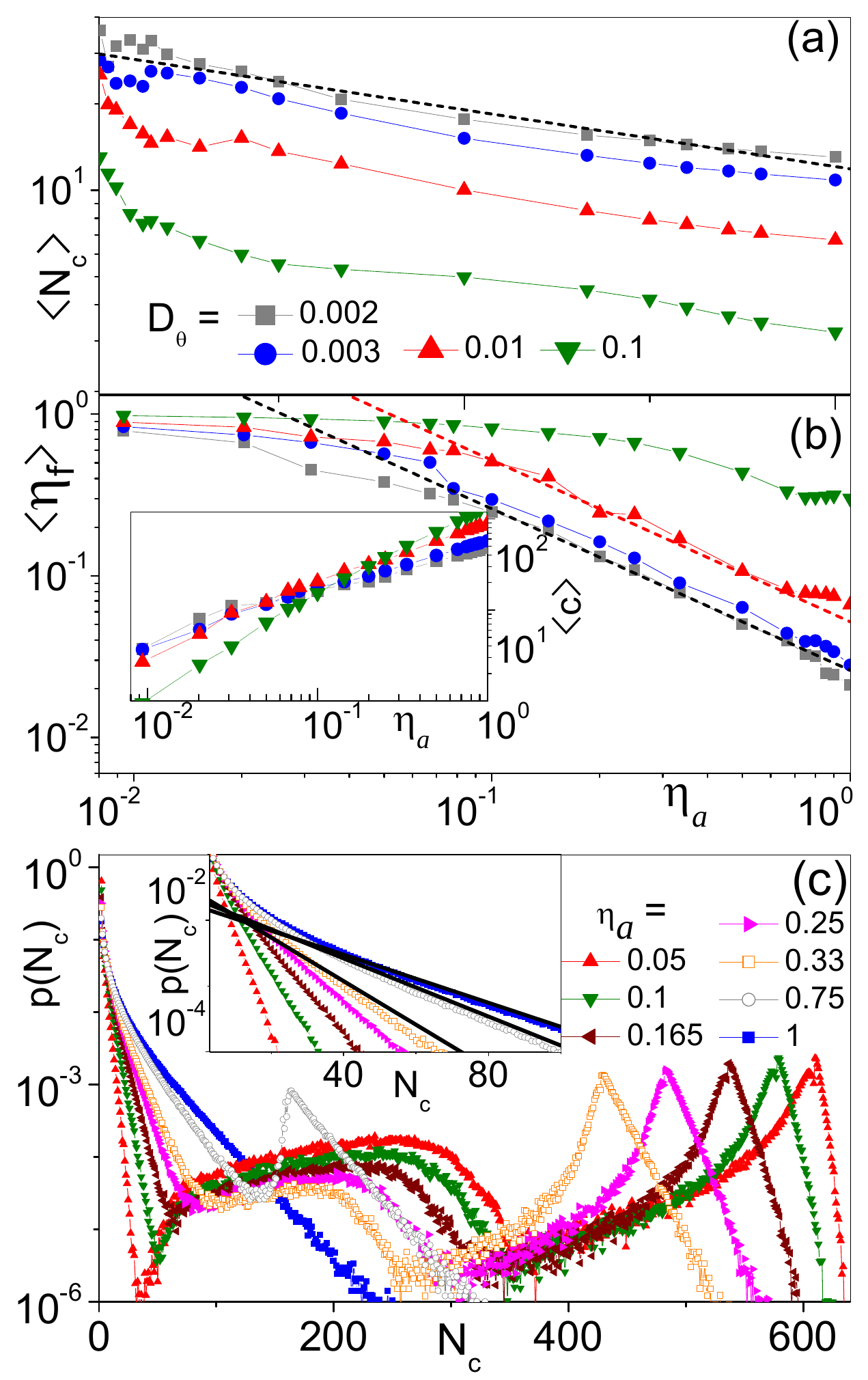}
\caption {(Color online) (a) $\langle N_c \rangle \;  vs.  \; \eta_a$ for different rotational diffusion constants $D_\theta$ (see legends) when active particles are added from one side of the file ({\it configuration-II}). The dashed line displays an eye guide for exponential decay.   (b) Similar figure as (a) but $ \langle \eta_f \rangle \;  vs.  \;\eta_a $. The dashed line represents the inverse relation between average fraction of free particles and active fraction in the mixture ($\langle \eta_f \rangle \sim 1/\eta_a$). Inset depicts average cluster numbers $\langle c \rangle$ as a function of $\eta_a$.  (c) Cluster size distribution with varying mixture composition (see legends). Inset is the distribution for small cluster size regime showing exponential decay tails.  The solid black line is the estimation of exponentially decaying tail based on the Eq.~(\ref{cluster-dis1}), with the substitution of $\rho_N$ by $\tilde{\rho}_N$ (see text).  Other model parameters are (unless reported otherwise in the legends): $ \; v_0 = 1.0, \;  \rho_N = 0.16,\; D_0 = 0,\; {\rm and} \; D_\theta = 0.002$. }
\end{figure}

With gradually adding active particles in the mixture, after a certain critical concentration, when separation between two adjacent active particles is close enough to interfere with each other's dynamics, the average cluster size decreases with $\eta_a$. As expected, this critical concentration is sensitive to the self-propulsion length. The transition point in Fig.~4(a) moves to the large $\eta_a$ in going from the low to high rotational diffusion. In this regime total passive particle sequences are equally divided by active pushers. Each segment behaves as a cluster with average size, 
\begin{eqnarray}\label{cluster-m1}
\langle N_c \rangle \sim N/N_a = 1/\eta_a.
\end{eqnarray}
This estimation [indicated by dashed line in Fig.~4(a)] well corroborates simulation results.

When active species concentration is high enough, the mixture behaves like an active single file.  In this region, both the cluster number and cluster size become independent of the $\eta_a$.  For $\eta_a \rightarrow 1$, one recovers Eq.~(\ref{cluster-1}). 

Figure 4(b) displays variation of the average fraction of the particles that survive as a single. It can be estimated as, 
\begin{eqnarray}\label{free-2}
\langle \eta_f \rangle = \frac{N-\langle c \rangle \langle N_c \rangle}{N}
\end{eqnarray}
Recall that in the limit $\eta_a \rightarrow 0$, the $\langle c \rangle$ is equal to the number of the active particle in the system. Further, $\langle N_c \rangle$ is independent of the mixture composition. This yields, 
\begin{eqnarray}\label{free-3}
\langle \eta_f \rangle  = 1- \eta_a \langle N_c \rangle
\end{eqnarray} 
Estimation based on the above equation [dotted line in the Fig.4(b)] accords with the simulation data. For quite a high active particle concentration $\langle \eta_f \rangle$ is very low and it exhibits a logarithmic decay with $\eta_a$.

{\it Cluster size distribution ---} To better understand the structure of the binary mixture we examine cluster size distribution, $p(N_c)$, with varying file composition. For $\eta_a \rightarrow 1$, $p(N_c)$   monotonically decays with $N_c$.  In the athermal condition, numerical results in Fig.4(c) witness fast (for small $N_c$) and slow (for large $N_c$) decaying segments in $p(N_c)$. The long tail decaying part is apparently exponential. However, the distribution function considerably deviates from,   $\left(1/\langle N_c \rangle \right) \exp\left[ -N_c/\langle N_c \rangle \right]$,  due to contribution of the initial fast decaying segment. The decaying tail of the distribution can be represented as, 
\begin{eqnarray}\label{cluster-dis1}
p(N_c) \sim \exp\left[-\frac{C N_c}{\sqrt{v_0\rho_N\tau_\theta}}\right]
\end{eqnarray}
The above equation is best fitted with simulation results [in Fig.4(c)] for $C=0.38$. With the introduction of passive particles noticeable changes in the size distribution are: (i) the tail of $p(N_c)$ decays faster, and (ii) an oscillating component around the decaying tail emerges. 

 With  decreasing $\eta_a$ the effective self-propulsion velocities of the active species get suppressed as they move pushing passive particles. It follows from Eq.(\ref{cluster-dis1}) that decaying tails become steeper with decreasing $v_0$.

 The periodic structure of the distribution becomes clearly noticeable when $\eta_a \leq 1/2$. For this mixture composition, each active swimmer encounters $(1-\eta_a)/\eta_a$ passive particles on their either side. As a result, an active particle aggregating passive ones forms a cluster of size, $N_c = (1-\eta_a)/\eta_a + 1=1/\eta_a$. 
\begin{figure}
\centering
\includegraphics[width=0.4\textwidth,height=0.5\textwidth]{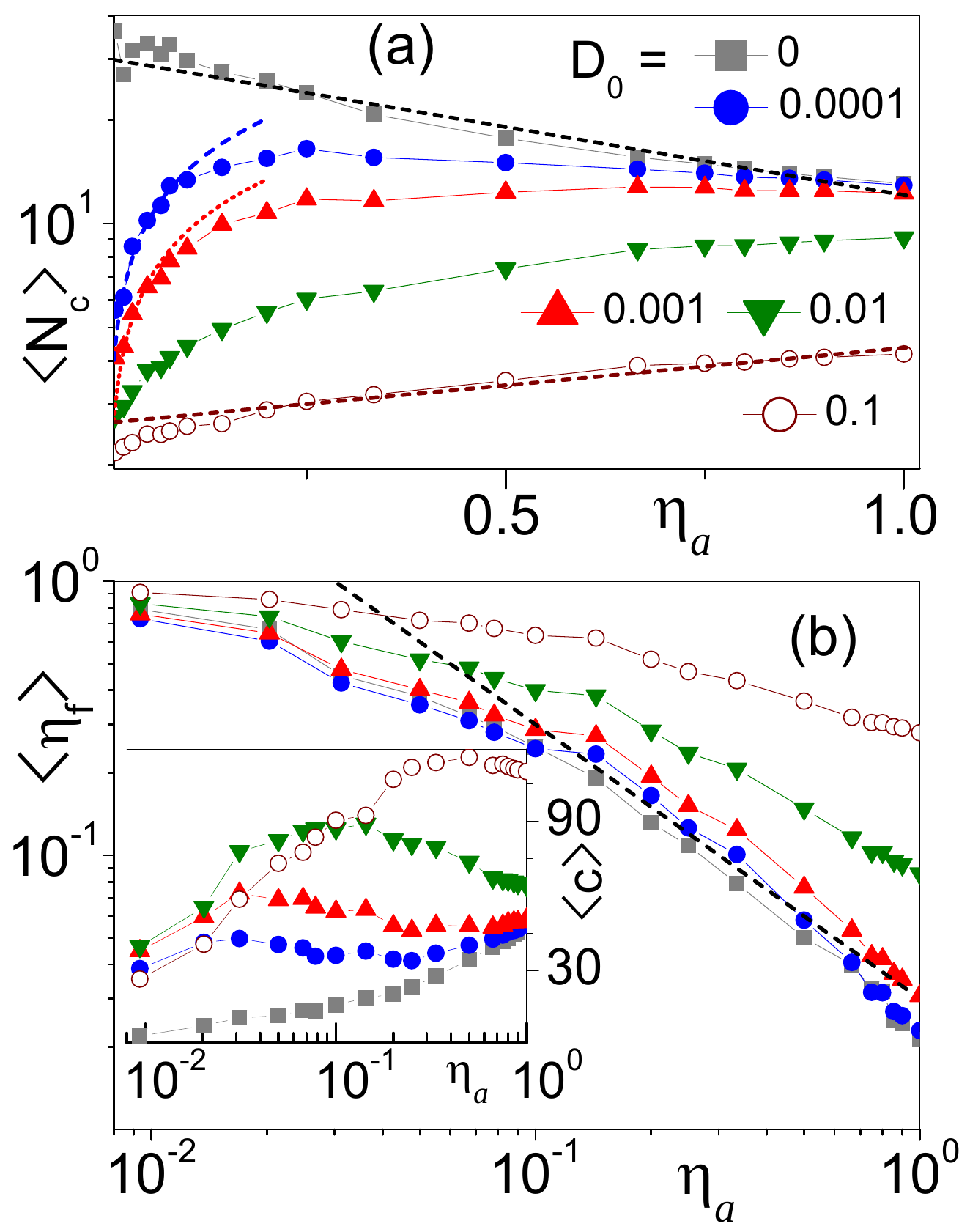}
\caption {(Color online) (a) $\langle N_c \rangle \;  vs.  \; \eta_a$  for different thermal diffusion constants $D_0$ (see legends) when active particles are added from one side of the file ({\it configuration-II}). The dashed line displays an eye guide for exponential decay.   (b) Similar figure as (a) but $ \langle \eta_f \rangle \;  vs.  \;\eta_a $. The dashed line represents the inverse relation between average fraction of free particles and active fraction in the mixture ($\langle \eta_f \rangle \sim 1/\eta_a$). Inset depicts average cluster numbers $\langle c \rangle$ as a function of $\eta_a$.    The remaining simulation parameters are (unless reported otherwise in the legends): $ \; v_0 = 1.0, \; \rho_N = 0.16, \; {\rm and} \; D_\theta = 0.002$. }
\end{figure}
Further, two or more such clusters can be aggregated to form larger clusters. This produces  peaks of $p(N_c)$  at $N_c = m/\eta_a$, with $m=1,2,3 ... $. When $\eta_a$ is neither too small (such that huge passive particles in between two swimmers can not be aggregated by pushing action ) nor too large, the height of the first peak is much larger than the other peaks. In this regime, the most probable peak position $N_c = 1/\eta_a$ coincides   with the average cluster size. This  estimation accords with Eq.~(\ref{cluster-m1}).

{\it Effects of the thermal fluctuations ---}  The presence of the thermal fluctuations leads to a considerable structural change of the file. It is manifested in the variation of $\langle c \rangle$, $\langle N_c \rangle$ and $ \langle \eta_f \rangle$ with changing the mixture composition. It is apparent from the Fig.~5, the thermal fluctuations break the large clusters into a number of smaller segments. Thus, for the very little fraction of the active particles in the mixture,  the clusters are  much smaller, however, they are more in number in comparison to the corresponding athermal limits. Nevertheless, similar to the athermal condition, the cluster size remains insensitive, and the cluster number linearly grows with  the increase  of active particle's concentration. 

Beyond the critical concentration, when an active particle starts encountering impact of self-propulsion of the adjacent particles, depending upon the strength of the thermal fluctuations, the $\langle c \rangle$  decays very slowly or remains unchanged with $\eta_a$. Further, the cluster size very slowly increases with increasing $\eta_a$. This is in sharp contrast to the athermal limit where $\langle c \rangle$ and $\langle N_c \rangle$ keep growing and decaying, respectively, with  increasing $\eta_a$  unless active particle concentration is too high.  For the very high concentration of the active particles, the mixture behaves alike both in the thermal and athermal limits. 

Figure 5(b) depicts that $ \langle \eta_f \rangle$ exhibits similar variation with $\eta_a$ as in the athermal situations unless the strength of the thermal fluctuations is too large. This attributes to the fact that the key quantity, $ \langle c \rangle \langle N_c \rangle $, which determines the value of $ \langle \eta_f \rangle$,  remains unchanged despite the splitting of the large clusters into smaller pieces due to thermal fluctuations.

\subsection*{Configuration II: Active particles are in the one side of the file}

We now consider the situation where active particles are added from one end of the file. Such situations may be encountered in cleaning micro-channels using self-propelled particles. In contrast to the {\it configuration I}, here two types of clusters, pure active and active-passive, can also be formed. Further, passive particles are pushed either by a single active or cluster of $n$ active particles with their centre of mass velocity $v_c=\sum_{i=1}^n v_{i}/n$. The velocity of the active cluster can be varied from its minimum, $v_c=0$, to the maximum $v_c=v_0$.  The cluster centre of mass velocity is a function of the self-propulsion parameters $D_\theta$ and $v_0$. Also, both the amplitude and persistence length of $v_c$ decay with the increasing cluster size. Thus, the size of the active-passive cluster decreases with increasing active particles in the system. Further, size of the active cluster can be estimated using Eq.~(\ref{cluster-1}). Again, with increasing $\eta_a$, {\it pushing effect} of active particles makes more free space for them. This leads to decrease of the active cluster size in the mixture. 

Based on the above reasoning, a decreasing trend for the average cluster size with $\eta_a$ is expected in the athermal condition. Simulation results presented in Fig.
6(a) show that $\langle N_c \rangle$ exponentially decays with $\eta_a$. The decay constant and the pre-exponential factor  depend on the self-propulsion parameters as well as $\rho_N$.  With increasing $\eta_a$, cluster number in the passive particle region does not change, however, in the active particle portion cluster numbers grow. Thus, the average cluster number almost linearly increases with adding active particles in the mixture.  
In the mixture most of the active particles exist as a single, as a result $\langle \eta_f \rangle$ remains unchanged with variation of the file composition unless  $\eta_a$ is large.

{\it Cluster size distribution ---}$p(N_c)$, shown in Fig.~6(c), supports our assertion about formation of pure active, as well as, active-passive clusters. The size distribution function first exponentially decays, then slowly increases and decays again forming a plateau like structure, finally, a sharp peak appears in the large cluster size regime. These three segments of $p(N_c)$ provide a clear picture about the mixture structure.  

 Similar to an active single-file, the first segment of the distribution exponentially decreases with $N_c$. 
The associated decay constant is inversely proportional to  $\sqrt{v_0\tau_\theta}$ (not shown in the figures). Further, the decay constant grows with decreasing $\eta_a$. These features are attributed to the formation of clusters with active species only. The active particles push passive ones and make more free space available for them. Thus, the effective number density in the active segment of the file is much smaller than the  passive part. If active swimmers are strong enough to aggregate passive species, effective density for the active part $\tilde{\rho}_N \sim \eta_a \rho_N/[1-(1-\eta_a)\sigma \rho_N]$. Thus, the exponentially decaying part of $p(N_c)$ could be described by the Eq.~(\ref{cluster-dis1}). However, $\rho_N $  there should be replaced by the effective density $\tilde{\rho}_N$. Further, this estimation requires correction in $v_0$ unless it is large enough to sustain impact of passive species. This analytic argument well corroborates simulation results in Fig.~6(c).

After the initial decay, the increasing trend in $p(N_c)$ witnesses formation of clusters due to the pushing effect of active particles. As expected, the probability of forming such clusters with larger size increases with increasing passive fractions in the mixture. As the active particles can aggregate a finite number of particles, after a certain cluster size $p(N_c)$ decreases sharply.   

The very sharp peak in $p(N_c)$ [at $N_c=N(1-\eta_a) $, see Fig.~6(c)] is due to finite size of the simulation box.  As a result of periodic boundary conditions, either end of the passive segment of the single-file encounters motile active species. Thus, all the passive particles can be aggregated forming a large cluster of size  $N(1-\eta_a) $.


The impact of the thermal fluctuations on the file structure is apparent from the Fig.7. Similar to the {\it Configuration I}, even very little thermal fluctuations break large clusters. With increasing $\eta_a$, effects of the thermal noise get suppressed when $v_0 \gg v_{th}$.

For the sake of simplicity we consider  active and passive alike, they are of the same size. Thus, they encounter the same damping and have similar thermal diffusivity in the unconstrained situations.  However, in general, active and passive species may not be of the same size. This would amount to the different thermal diffusivity and damping strength.  A smaller passive  particle encounters less damping, as a result, it can move more easily due to motility transfer from the motile swimmers.   When a passive particle is smaller (larger) than active ones, more (less) number of passive particles can be aggregated by the pushing action of active species. This effect is equivalent to the increasing (decreasing) strength of self-propulsion. 

\section{Diffusion} 
\begin{figure}
\centering
\includegraphics[width=0.45\textwidth,height=0.32\textwidth]{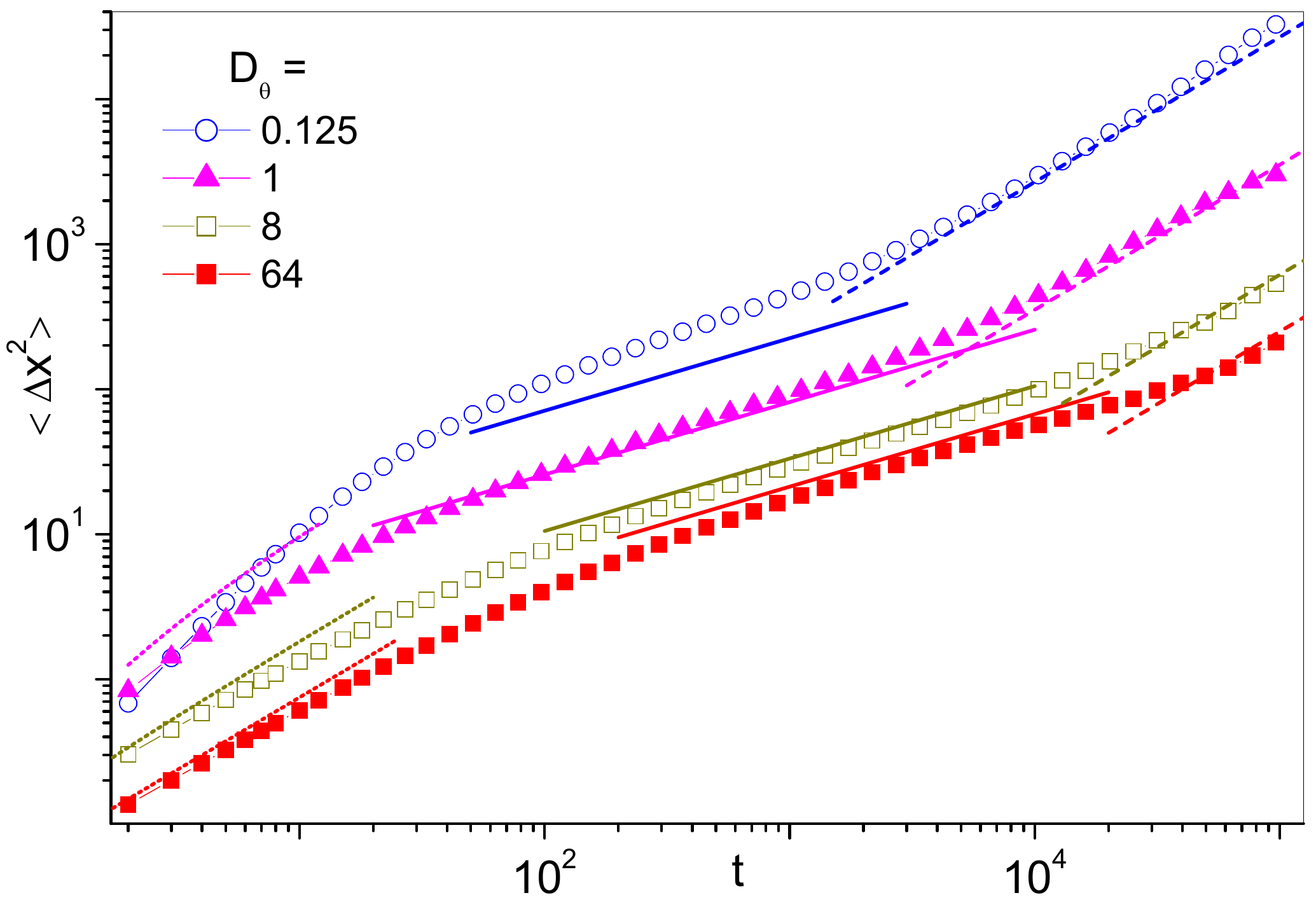}
\caption {(Color online) Mean square displacement  $\langle \Delta x^2  \rangle$ as a function of time for different rotational diffusion constants (see legends).  Three regimes of diffusion are fitted with analytic formulas. The dotted lines depict mean square displacement for the short regime based on the Eq.~(\ref{dif2}). Solid lines represent analytic estimation of sub-diffusion regime based on the Eq.~(\ref{t-half1}). The long time normal diffusion regime is fitted with the formula $\langle \Delta x^2  \rangle = 2D_s t/N$. Other simulation parameters (unless reported otherwise in the legends): $ \; v_0 = 1.0, \; \rho_N = 0.1, \; {\rm and} \; D_0 = 0.03$.}
\end{figure}

To understand diffusion of self-propelled particles as well as active-passive binary mixture as a single file, we calculate mean square displacement,
\begin{eqnarray}
 \langle \Delta x^2  \rangle = \sum_{i=1}^N \langle \left[ x_i(t) - x_i(0)\right]^2 \rangle  \label{dif1}
\end{eqnarray}
$\langle ....  \rangle $ denotes averaging over a large number of independent stochastic realization. The mean square displacement $\langle \Delta x^2 \rangle$ as a function of $t$ is plotted in Fig.8 for different rotational diffusion $D_{\theta}$. 
Three distinct regions are apparently distinguished here. Initially, free diffusion of Janus particles (JPs) is observed with mean square displacement, 
\begin{eqnarray}
 \langle \Delta x^2  \rangle = 2\left(D_0 + \frac{v_0^2}{2D_{\theta}}\right)t+ \frac{2v_0^2}{D_{\theta}^2}  \left( e^{-D_{\theta}t}- 1 \right)   \label{dif2}
\end{eqnarray}
As soon as particles meet neighboring ones, diffusion occurs under the constraint, particles cannot cross each other. Thus, a transition in $\langle \Delta x^2 \rangle$ versus $t$ is observed after a certain time $t_c$ which is the  measure of the  time to cross the average inter particle distance $L/N$.  The transition time $t_c$  depends on the particle density, particle size and diffusion of free Janus particles. 

For $t>t_c$, the mean square displacement displays a sub-diffusive behavior, 
\begin{eqnarray}
 \langle \Delta x^2  \rangle = \frac{2(1-2r_0\rho_N)F}{\rho_N} \,t^{1/2}  \label{t-half1}
\end{eqnarray}
  The mobility factor $F$ is related to the single Janus particle diffusion constant $D_s$ as, $F = \sqrt{D_s/\pi}$. The diffusion constant of a free Janus particle is given by,
 \begin{eqnarray}
D_s = D_0 + {v_0^2}/{2D_{\theta}}  \label{Ds}
\end{eqnarray}

The diffusion constant consists of two parts. The first term is the diffusion contribution due to thermal fluctuations.  The pure thermal diffusion contribution can be computed by measuring the translational diffusion of a free JP in the absence of self-propulsion. The second part in Eq.(\ref{Ds}) is the contribution of self-propulsion motion, directly proportional to the persistence time $\tau_\theta$ and  modulus of self-propulsion velocity $v_0$.

\begin{figure}
\centering
\includegraphics[width=0.45\textwidth,height=0.32\textwidth]{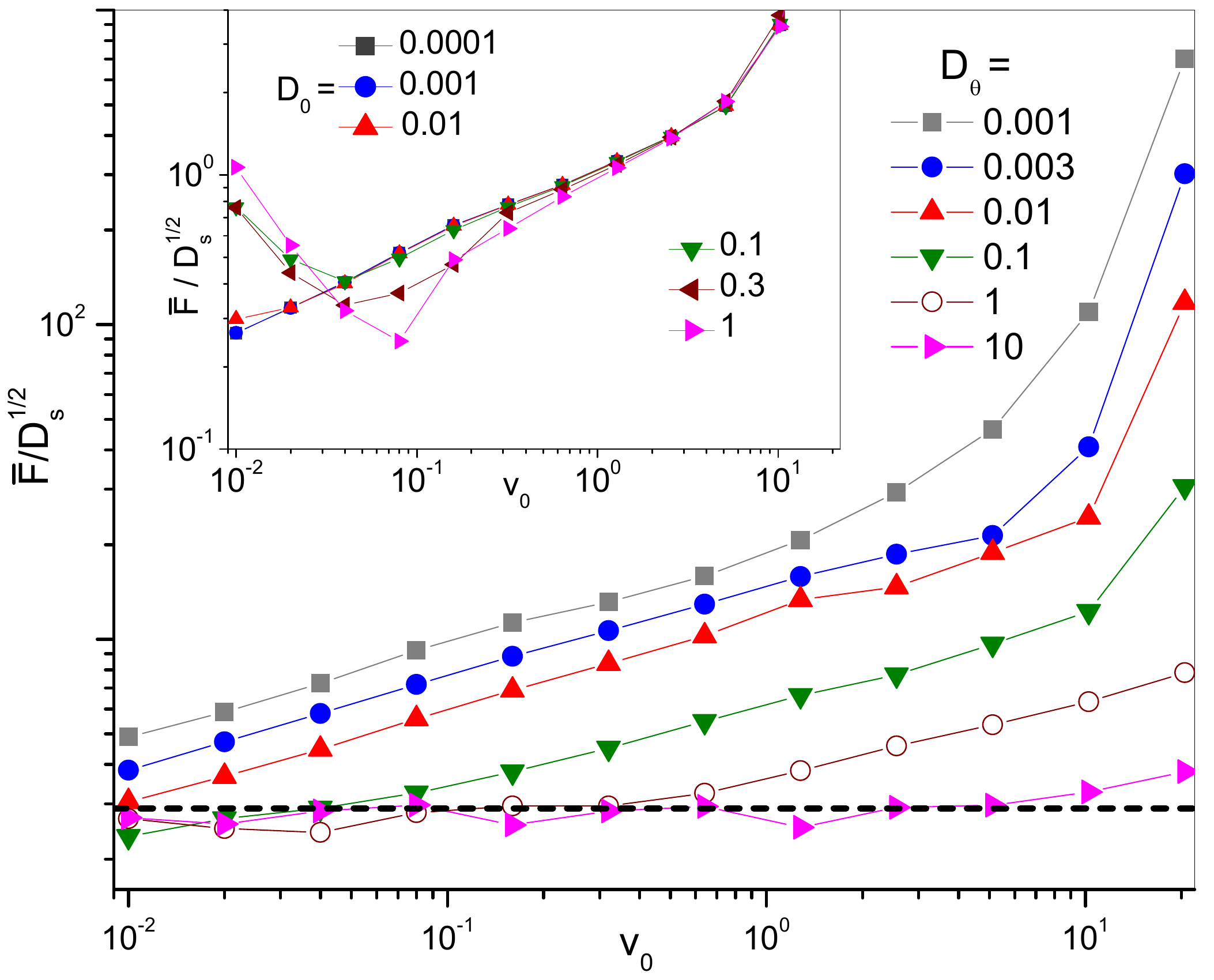}
\caption {(Color online) $\overline{F}/\sqrt{D_s}$ versus $v_0$ for different rotational diffusion (main panel)  and the thermal translational diffusion (inset)  as shown in the legends. Dashed line is the estimation based on Eq.~(\ref{mobi-thermal}). Simulation parameters unless mentioned in the legends: $\rho_N=0.16, \; D_0 = 0, \; {\rm and} \; D_\theta = 0.002$.}
\end{figure}

In the Fig.8, the theoretical estimates of  the mean square displacement in the normal diffusion and sub-diffusion regimes are depicted by dotted and solid lines, respectively. For normal diffusion region, simulation results are in good agreement with analytic ones for all parameters. Also, in the sub-diffusion regime,  analytic estimate based on the Eq.(\ref{t-half1}-\ref{Ds}) is in accord with simulation for fast rotational diffusion of the JPs.  However, the disagreement becomes noticeable  for large self-propulsion persistence $\tau_{\theta}$. This attributes to the interference between stochastic and ballistic single file diffusion.

When the time is long enough for the tag particle to diffuse the length $L$ normal diffusion region is recovered. This regime  describes diffusion of the entire file as a whole. As if particles are strongly coupled and we are observing diffusion of the file's center of mass. Thus, mean square displacement is given by, $ \langle \Delta x^2  \rangle = 2D_s t /N$.  

\subsection{Diffusion in a single file of active particles} 
For the all self-propulsion parameter regimes, in the long time limit and for very large $L$, the mean square displacement follows the relation, 
\begin{eqnarray}
\langle \Delta x^2 \rangle = 2 \overline{F} t^{1/2}  \label{t-half2}.
\end{eqnarray}

Where, $\overline{F}$ is a function of the self-propulsion parameters in addition to the particle density $\rho_N$ and intrinsic thermal transnational diffusion $D_0$. In the Fig.~9 we plot the mobility $\overline{F}$ (in the unit of $\sqrt{D_s}$ ) as a function of self-propulsion velocity for different rotational diffusion constant under athermal condition. The inset here represents impacts of $D_0$ in $\overline{F}/\sqrt{D_s}$ versus $v_0$.

For a self-propulsion length larger than the average inter particle spacing, active particles preferably exist as clusters of different sizes. Simulation results in Fig.~9 show that for the diffusion of a tag particle buried in the cluster, the mobility is proportional to the $v_0^\alpha/D_\theta^{\beta_3}$ as long as $v_0 < v_0^c$. Where the exponents, $\alpha > 1$ and $\beta_3 > 0.5$. However, in the opposite limit, when the self-propulsion length is much smaller than the inter-particle spacing, particles in the 1D system preferably diffuse as a single particle with the constraint of single file.  Particle dynamics here can be assumed as similar to the passive particle with effective temperature, 
\begin{eqnarray}
T_{eff} = \frac{\gamma}{k_B}\left( D_0 + \frac{v_0^2}{2D_{\theta}} \right)  \label{Teff}.
\end{eqnarray}

Thus, under athermal condition the mobility factor becomes,
\begin{eqnarray}
 \overline{F} \rightarrow \frac{1-\rho_N}{\sqrt\pi \rho_N} \label{mobi-thermal}.
\end{eqnarray} 
\begin{figure}
\centering
\includegraphics[width=0.45\textwidth,height=0.6\textwidth]{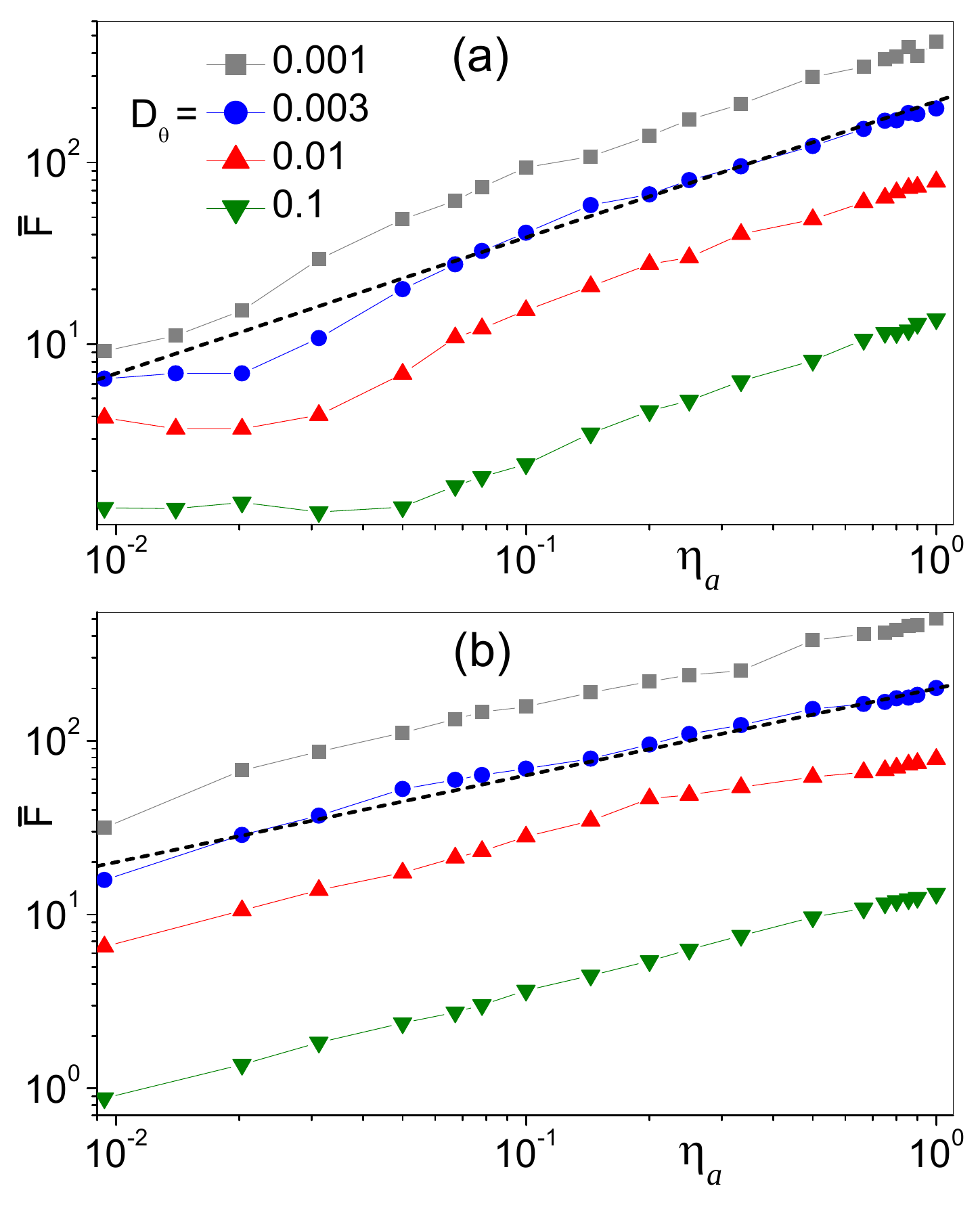}
\caption {(Color online) $\overline{F}$ versus $\eta_a$ for different rotational diffusion for two types of file configurations: (a) {\it Configuration I}: Active particles are uniformly distributed throughout the mixture; (b) {\it Configuration II}: Active particles are in the one side of the file and the other side contains passive particles only. The dashed line represents best fitted empirical relation, $\overline{F}=\overline{F}_0 \eta_a^{\kappa} $, with $\kappa = 0.75$ and $0.5$, respectively, for the configurations I and II.  Simulation parameters unless mentioned in the legends: $\rho_N=0.16, \; D_0 = 0, \; {\rm and} \; v_0 = 1.0$.                                                                                                                                                                                                                                                                                                                                                                                                                                                                                           }
\end{figure}
 This estimation (indicated by dashed line in Fig.~9) is in well accord with simulation data when $l_\theta \ll 1/\rho_N$. 

The inset in Fig.~9, depicts impact of thermal fluctuations on diffusion of active single file. Here, we consider the parameter regime with quite a slow rotational relaxation time so that movements are correlated for the entire range $v_0$. Thermal fluctuation impact becomes noticeable when $D_0 > {v_0^2}/{2D_{\theta}}$. Note that for both in the thermal and athermal conditions, the mobility starts growing much faster  with $v_0$ as soon as the self-propulsion velocity gets close to its critical value $v_0^c$. Recall that beyond the critical value, self-propulsion force is strong enough to make particles overlap each-other overcoming strong repulsion. It makes more space for free movement of the particles.

\subsection{Diffusion in a single file of active-passive binary mixture}
Simulation results  in Fig.~10(a,b) depict diffusion of a tagged particle in the active-passive binary mixture.  It should be noted that diffusion of a tagged particle, whatever it is active or passive one, displays similar traits. In the both types of configurations of active passive mixture, the diffusion of a tagged particle is greatly enhanced with increasing fraction of active particle in the mixture. This attributes to the motility transfer of motile active particles to slowly diffusing passive particles\cite{motility-transfer}.  

For {\it configuration I}, where active particles are uniformly distributed throughout the mixture, initially with adding active particles the mobility $\overline{F}$ is almost insensitive to $\eta_a$ [see Fig.~10(a)]. Increasing the active fraction in the binary mixture beyond a certain value, mobility enhances monotonically. For quite large $\eta_a$, simulation data are best fitted with the empirical relation, $\overline{F}=\overline{F}_0 \eta_a^{0.75} $. It is depicted by a dashed line in Fig.~10(a).  

On the other hand, when active particles are added from the one side to a single file of passive particles ({\it configuration II}), the mobility is a monotonically increasing function of $\eta_a$ for the entire range of
mixture composition. Best fitted empirical relation, $\overline{F}=\overline{F}_0 \eta_a^{0.5} $ [see Fig.~10(b)]. For the both configurations of the binary mixture, $\overline{F}_0$ is a function of self-propulsion parameters, $D_\theta$ and $v_0$. Further, in the both configuration diffusion of the file considerably enhances through motility transfer mechanisms.  

\section{Conclusions} 
We have investigated the structural and diffusion properties of a mixture of
active and passive colloidal particles of finite size in a single-file. The combination of self-propulsion and steric collisions results peculiar clustering in one dimension both in the pure active particle system, as well as, in the binary mixture.  Our simulation results show that for the vanishingly small thermal fluctuations, when the self-propulsion length is larger than the average interparticle separation, active particles in a single file exist as  clusters of various sizes. The average cluster size is a monotonic function of the self-propulsion length.

In the binary mixture a very similar clustering property is observed for  {\it configuration I} when the fraction of active particles in the system is more than 10\%. For very low active particle concentration, each active tracer forms clusters aggregating passive ones through pushing action. In this regime cluster size is insensitive to  active species concentration. On the other hand, when active swimmers are added from one side of the mixture ({\it configuration II}), a much smaller number of   clusters are formed  and their size decay exponentially with increasing fraction of active particles in the mixture. For the both mixture configurations, cluster size is very sensitive to the thermal fluctuations even to a small extent.

Tagged particle diffusion in the single file of binary mixture exhibits sub-diffusive behaviour with $\langle \Delta x^2 \rangle = 2\overline{F}t^{1/2}$. Where, the mobility factor $\overline{F}$ monotonically grows with increasing concentration of active particles in the mixture.   This enhanced diffusion in the active-passive binary mixture witnesses motility transfer from the energetic active particles to the slowly moving passive ones.     

We hope all our simulation results
would be useful to synthesise active Janus particles with desired transport
features and to operate them in a controlled manner for
novel and emerging applications. In particular, to mention, targeted drug delivery, cleaning biological channels, and other emerging  nano-technological and biomedical applications where particles diffuse through narrow channels with non-passing constraints.

\section*{Conflict of Interest}
The authors declare no conflict of interest.

\section*{Data Availability Statement}
The data that support the findings of this study are available on
request from the corresponding author. The data are not
publicly available due to privacy or ethical restrictions.

\section*{Accknowledgement}
P.B. thanks UGC, New Delhi, India, for the award of a Junior
Research Fellowship. P.K.G. is supported by SERB Core Research Grant No. CRG/2021/007394.

\end{document}